 \documentclass[prl,a4paper,
twocolumn,amsmath, amssymb, floatfix,
nofootinbib,wrapfloat byrevtex]{revtex4}

 \usepackage{amsmath,mathrsfs,upgreek,graphicx,placeins,float}
 \usepackage{color}
 \usepackage{booktabs}
 \usepackage{multirow}

 \newcommand{\beq}[1]{\begin{equation}\label{#1}}
 \newcommand{\eeq}{\end{equation}}
 \newcommand{\bea}[1]{\begin{eqnarray}\label{#1}}
 \newcommand{\eea}{\end{eqnarray}}
  \newcommand{\bec}[1]{\begin{center}\label{#1}}
 \newcommand{\eec}{\end{center}} 
 \newcommand{\mm}{{\rule[1.5pt]{3pt}{0.6pt}}}

\begin{document}

 \title{Complementarity of Gravitational Collapse \\ (I) Origin of the Bekenstein-Hawking Entropy}
 \author{Ding-fang Zeng}
 \email{dfzeng@bjut.edu.cn, ORCID: https://orcid.org/my-orcid?orcid=0000-0001-5430-0015}
 \affiliation{Beijing University of Technology, Chaoyang, Bejing 100124, P.R. China,
 }

\begin{abstract}
Through two exact solution families to the Einstein equation and one-to-one correspondence between their free parameters, we show that the ensemble of collapsars with arbitrarily close-to-implementing horizon in the Schwarzschild time definition and the over-cross-oscillatory solid-ball ergodically experiencing all possible modes in the Lema\^itre time definition constitute two complementary description for the inner structure of black holes formed through gravitational collapse. As a support for this complementarity, we prove that the area law formula of Bekenstein-Hawking entropy by counting the degeneracy of collapsing material's wave functional directly. In two companion works, observational signals of this inner structure picture will be reported independently.
\end{abstract}
 \maketitle
  
{\em Introduction} What's the form of matters inside physical black holes (pBHs), i.e. those formed through gravitational collapse, is a fundamental question of modern theoretical physics and astronomy. The popular answer to this question is based on a series of singularity theorem proved by Penrose, Hawking and Geroch et al \cite{Penrose1965,Hawking1976,geroch1979}. According to their theorems, in spacetime controlled by general relativity (GR) with suitable energy condition, the formation of trapped surface inevitably leads to the existence of finite affine length light-like (FALL) geodesics, thus making pBHs coincide with those solved from Einstein equation with singular sources. However, with concrete examples in 2023 \cite{Kerr2023}, Roy Kerr argued that FALLs do not necessarily terminate on non-extendable point of spacetime, so the statement that gravitational collapse must cause incompleteness or unpredictability is a belief rather than logic. Nevertheless, the prevailing view remains that by GR the only form of matter's existence inside pBHs is singularity, i.e., points or circular-lines with infinite mass-energy density and spacetime curvature. At the same time, by arguing that such singularities form always behind event horizons so are undetectable to all outside probes \cite{CCH1969,CCH1979}, this belief puts itself on a position unfalsifiable.

However, this belief contradicts the basic fact of BH thermodynamics directly, especially that BHs are entropic so microscopically diversifying. Facing this contradiction, both string theory \cite{strominger1996,HorowitzStrominger9602,fuzzballsMathur2002,fuzzballSkenderis2007} and loop quantum gravity \cite{lqgEntropy1996,lqgEntropy1997,AshtekarBojowald0509,Modesto0509} choose giving up the singularity belief but embrace thermodynamics differently; while some classic \cite{softhairHPS2016} and semi-classic magicians \cite{vijay2024prl} try dancing with the two simultaneously under musics inaudible to classic gravitational probes. Except the fuzzy ball picture of \cite{fuzzballsMathur2002,fuzzballSkenderis2007} which can be considered a string theory implementation of our picture here, inner structure of BHs in all these theories is undistinguishable from that of singularity belief by the current and near future observations.  In contrast with these works, we will report here a new reconciliation between the BH thermodynamics and singularity theorem, whose inner structure picture for pBHs \cite{dfzeng2023} is easily dis-verifiable. Our key point is the general coordinate invariance of gravitational physics, according to which the microscopic state of pBHs can be equally well described in both the Schwarzschild and Lema\^itre time definitions. 

Our logic goes as follows. (i) in the Schwarzschild time definition, the gravitation collapse of a massive star causes only  a continuously contracting collapsar with close-to-implementing but never successfully implemented event horizon; dynamic perturbation and relaxation will assign the collapsar an approximately $m(r)\stackrel{t{\rightarrow}\infty}{=}r/2G_{\scriptstyle N}$-type radial mass profile, in spherically symmetric cases e.g.; an ensemble of collapsars with the same symmetry and total mass is needed to account for the initial configuration's uncertainty. (ii) in the Lema\^itre time definition, the formation singularity is not the terminal of physical evolution; matters hitting on the singularity over-crosses, so both the singularity and horizon forms and resolves periodically. Unpredictability of physics across the singularity will make oscillations of the system ergodic. (iii) the Ensemble of collapsars in the Schwarzschild time definition (EnS) and the Ergodically over-cross-oscillatory modes of the system in the Lema\^itre time definition (ErL) form two equally well and complete descriptions for the microscopic state of pBHs, i.e. EnS=ErL.

We will call this relationship of equivalence and completeness as Complementarity of Gravitational Collapse (CGC). Comparing with the BH complementarity principle of refs. \cite{complementarity1990thooft,complementarity1993schoutens,complementarity1993}, CGC is not a principle customised for solving the information missing puzzle, although it has such potentials \cite{dfzeng2023}; it is a general feature of gravitational physics derivable from the general covariance of GR. In our case, this only means that EnS can be obtained from the modes of ErL through the time coordinate's redefinition. On the developments of this recognition, we recommend readers to our eaerly works \cite{dfzeng2017,dfzeng2018a,dfzeng2018b,dfzeng2020,dfzeng2021,dfzeng2022} and \cite{Stojkovic2008b,Stojkovic2014}. The observational evidence and predictions of this complementarity is rather remarkable and will be discussed in two other works \cite{dfzeng2025b,dfzeng2025c} independently. In the remaining part of this work, we elaborate details of this complementarity, from exact solution families to the Einstein equation in two time definitions to the proof of the area law formula of Bekenstein-Hawking entropy through quantization, and to discussions about potential observational evidences.

{\em Exact solution families} In the case of the gravitational force dominates over all other interactions, neglecting the material pressure of a collapsar is reasonable for the purpose of its microscopic state counting. So to avoid splitting attention to non-gravitational physics, we will focus on the collapse of spherical dust balls in this work only, leaving studies of the astronomically interesting stellar's collapse to ref.\cite{dfzeng2025c}. By the Schwarzschild time definition, it can be proven \cite{dfzeng2023,dfzeng2017} that the full space metric of such a collapsar family can be written as
\bea{}
&&\hspace{-5mm}ds^2_\mathrm{full}{=}-{e^\Phi}dt^2{+}h^{-1}dr^2{+}r^2d\Omega^2, h{=}1{-}\frac{2GM[t,r]}{r},
\label{metricOFO}
\eea
The instantaneous Newtonian potential $\Phi$ contains ambiguity related with the definition of $t$. This ambiguity can be eliminated by requiring that the collapsing rate of $r$-shell at time $t$ equal to the freely falling velocity of test-particles $r$-away from a Schwarzschild-BH of mass $m(t,r)$, that is
\beq{}
\noindent{}v(t,r)^{radial~contraction}_{collapsing~material}=v(r)^{freely~falling~partciles}_{outside~a~Schwz~m(t,r)}.
\label{schwzGaugeCondition}
\eeq 
The latter is derivable from the equation of motion of test particles outside a Schwz-BH of mass $m(t,r)$
\beq{}
\Big\{\begin{array}{l}h\dot{t}=\gamma_m\\h\dot{t}^2-h^{-1}\dot{r}^2=1\end{array}
\Rightarrow{}v=\frac{dr}{dt}=-\sqrt{(1-\gamma_m^{{-}2}h)h^2}
\label{vExpression}
\eeq 
where $\gamma_m=\sqrt{1-2m(0,r)/r}$ so that the initial speed is zero. This gauge choice would not trivialize physics of the collapsar, but assure all physics we are discussing happens in the coverage of time definition outside probes were selected and monitored.

Denoting the 4-velocity of collapsing volume elements as $u^\mu=\{u^t,u^tv,0,0\}$, then the Einstein equation $G^{\mu\nu}=8\pi{}G_{\scriptscriptstyle N}{\varepsilon}u^\mu u^\nu$ and the normalization of 4-velocity $u^\mu$ can be combined together to yield (we choose unit $8\pi G_{\scriptscriptstyle N}\equiv1$)
\bea{}
&&\hspace{-5mm}\varepsilon=\frac{2m'}{r^2}\frac{v^2-e^{\Phi}h}{e^{\Phi }h},u^tu^t=\frac{h}{v^2-e^{\Phi }h}
\label{epExpression}
\\
&&\hspace{-5mm}\dot{m}=-m'v,
\label{m10Expression}
\\
&&\hspace{-5mm}
\Phi'=\frac{2v^2 rm'{+}2 e^{\Phi }hm}{
r^3e^{\Phi }h^2}
\label{F01Expression}
\\
&&\hspace{-5mm}
\dot{\Phi}=-\frac{e^{-\Phi } m'v^3 }{rh^2}
-\frac{3mv}{hr^2}
+v'+\frac{\dot{v}}{v}
+
\frac{e^{\Phi }m}{r^2 v}
\label{F10Expression}
\eea{}
where $\dot{X}\equiv\frac{\partial X}{\partial t}$, $X'\equiv\frac{\partial X}{\partial r}$. Eqs.\eqref{m10Expression}\&\eqref{F10Expression} form a first order dynamic system with constraint \eqref{F01Expression}. This is a typical feature of Einstein equation. We checked that when these equations are satisfied, the conservation law $T^{\mu\nu}_{~;\nu}=0$ satisfies automatically. As a first order constraint, eq.\eqref{F01Expression} can be integrated from the boundary surface of matter occupation region, where  $e^\Phi(t,r_\mathrm{boundar})$ can be set to $1-\frac{2GM}{r}$ to join with the outside Schwarzschild smoothly. So, initial conditions for eqs.\eqref{m10Expression}\&\eqref{F10Expression} consists of $m(0,r)$ only. As results, the microscopic state of the collapsar is determined by $m(0,r)$ exclusively.

\begin{figure}[t]
\includegraphics[totalheight=80mm]{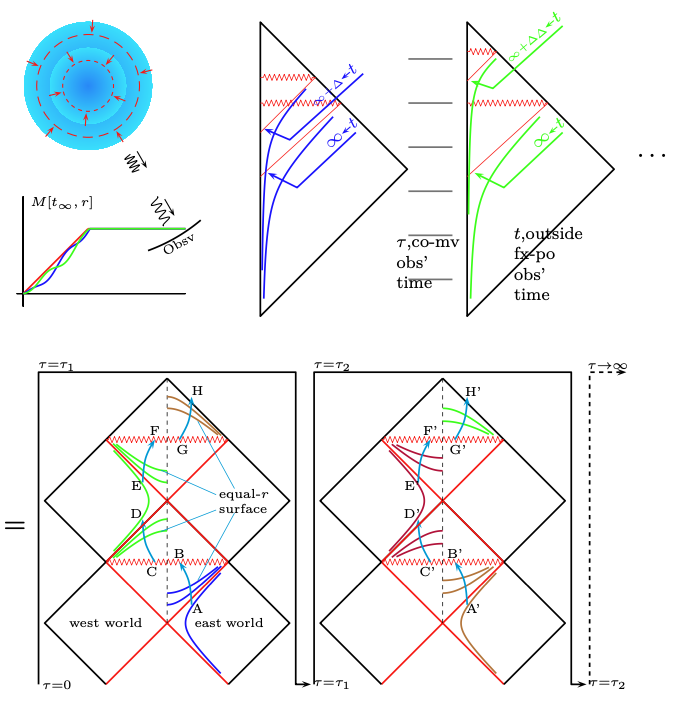}
\caption{In the Schwarzschild time definition, the horizon of a collapsar is only close-to-implementing but never successfully implemented; for any macroscopic collapsar, an ensemble of microscopic collapsars with equal mass and symmetry but different radial distribution is needed to account for the initial profile's multiplicity. In the Lema\^itre time definition, the central singularity is inevitable but it is not the terminal of physical evolution. The unpredictability of physics across the singularity only means that the collapsing material's oscillation across the central point is ergodic. This two pictures are equivalent due to general coordinate invariance.} 
\label{figTwoDescription}
\end{figure}

Both numeric and formula analysis can be used to show that $m(t,r)\xrightarrow{t\rightarrow\infty}\frac{r}{2}$. Due to the gravitational time dilation, to all probes defined in the Schwarzschild time definition, i.e. those selected and monitored in $-\infty{<}t{<}\infty$, the collapsar only manifests as a continuously contracting object with $m[\infty,r]=\frac{r}{2}+\mathrm{devi.}$ type radial radial mass profile. An ensemble of members with the same spherical symmetry but different radial mass-profile is needed to account for the deviation's multiplicity, see the upper part of FIG.\ref{figTwoDescription}. So even though the collapsar develops no successfully implemented horizon, it is entropic. In the next section we will prove that it is just this abandoned information related with the mass profile's deviation from the exact $m(r)=\frac{r}{2G}$-line that forms the basis of Bek-Hwk entropy. But before that proof, we need show that the Schwarzschild time is not a wrongly chosen time coordinate. In the Lema\^itre time definition, a totally peer-to-peer picture exists also.

Lema\^itre time is the proper time of probes co-moving with the collapsing material. By this time definition, the spacetime metric \cite{dfzeng2023,dfzeng2020} of dust collapsar with inhomogeneous radial mass profile can be written as
\bea{}
&&\hspace{-5mm}ds^2_\mathrm{in}{=}{-}d\tau^2{+}\frac{\big[1{-}\big(\frac{2m}{\varrho^3}\big)\!^\frac{1}{2}\!\frac{m'\!\varrho}{2m}\tau\big]^2\!d\varrho^2}{a[\tau,\varrho]}{+}a[\tau,\varrho]^2\varrho^2d\Omega^2_2,
\label{metricICO}
\\
&&\hspace{-5mm}ds^2_\mathrm{out}{=}{-}d\tau^2{+}\frac{r_s^{2/3}d\varrho^2}{[\frac{3}{2}(\varrho{-}\tau)^\frac{2}{3}]}{+}[\frac{3}{2}(\varrho{-}\tau)^\frac{2}{3}]^2r_s^\frac{2}{3}d\Omega_2^2
\label{outMetric},
\eea
\bea{}
&&\hspace{-5mm}a[\tau\!\in\!|_0^{\!\frac{p\;\!\!^\varrho}{4}},\varrho]{=}a_0\big(1\!-\!\frac{4\tau}{p^\varrho}\big)^{\!\frac{2}{3}}
,~a[\tau|_{\frac{p\;\!\!^\varrho}{4}}^{\frac{p\;\!\!^\varrho}{2}},\varrho]{=}{-}a[\frac{p\;\!\!^\varrho}{2}{-}\tau,\varrho],
\label{aperiodic}
\\
&&\hspace{-5mm}a[\tau|_{{p\;\!\!^\varrho\!/2}}^{p\;\!\!^\varrho},\varrho]{=}{-}a[p^\varrho{-}\tau,\varrho]
,~
a[\tau|_{p\;\!\!^\varrho}^{p\;\!\!^\varrho{\!\scriptscriptstyle+}},\varrho]=a[\tau{-}{p\;\!\!^\varrho},\varrho],
\nonumber
\eea
where $m=2GM[\varrho]\equiv{}m^{\scriptscriptstyle\!L}$ is the collapsar mass function on the co-moving grid $\{\varrho,\theta,\phi\}$ determined by initial conditions, while $a[\tau,\varrho]$ is an oscillatory function of $\tau$ with $\varrho$ dependent period $p^\varrho\equiv\frac{8}{3}\big(\frac{\varrho^3}{2GM[\varrho]}\big)\!^\frac{1}{2}$. The function form \eqref{aperiodic} follows from the Einstein equation in this time definition $R_{\mu\nu}-\frac{1}{2}g_{\mu\nu}R=\{\frac{m'/\varrho^2}{a^\frac{3}{2}-\frac{\tau\varrho m'}{2\sqrt{m\varrho^3}}+\frac{3\tau^2m'}{4\varrho^2}}(\equiv\!\varepsilon^{\scriptscriptstyle\!L}),0,0,0\}$. If one set $a_0=2^{-\frac{2}{3}}$ and $m^{\scriptscriptstyle\!L}[\varrho]=m^{\scriptscriptstyle\!S}[0,r]$, then $\varepsilon^L$ will reduce to $\varepsilon^S$ with $\dot{m}^{\scriptscriptstyle\!S}=0$ directly. Outside the matter occupation region, by the following coordinate transformation
\beq{}
dt{+}\sqrt{\frac{r_s}{r}}\big(1{-}\frac{r_s}{r}\big)^{\!-1}\!dr{\equiv}d\tau
,
dt{+}\sqrt{\frac{r}{r_s}}\big(1{-}\frac{r_s}{r}\big)^{\!-1}\!dr{\equiv}d\varrho,
\eeq
the metric \eqref{outMetric} will reduce to the outside part of metric \eqref{metricOFO} directly. It can be checked that all initial mass profile allowed by \eqref{metricICO} have correspondences in \eqref{metricOFO}, and vice versa. So, if the Lema\^itre time geometry completely describes the inner-structure of a BH with periodically forming and resolving horizon and singularity, then the Schwarzschild time geometry will do equally well.

The Lema\^itre time geometry \eqref{metricICO} contains singularity, which happens in finite duration after the collapse begins. However, such a singularity is not the terminal of physics evolution, the motion and oscillation of materials across that point is meaningful. In ref. \cite{CPTsymmUniverse2018,CPTsymmUniverse2022} Niel Turok discussed cosmological implications of a possible continuation. The unpredictability of physics across the singularity only means that such oscillations are ergodic. That is,  the full oscillation of collapsing material will experience all possible modes characterized by $m^{\scriptscriptstyle\!L}[\varrho]$. For example, before the singularity-crossing $\tau{\in}(-\frac{p^\varrho}{4},\frac{p^\varrho}{4})$, the metric \eqref{metricICO} has $m^{\scriptscriptstyle\!L}_1[\varrho]$, but after the singularity-crossing $\tau{\in}(\frac{p^\varrho}{4},\frac{3p^\varrho}{4})$, it has $m^{\scriptscriptstyle\!L}_2[\varrho]$. What was written in \eqref{aperiodic} is an oscillatory mode which crosses the singularity symmetrically. In the Schwarzschild time definition, this mode space will be characterised by $m^{\scriptscriptstyle\!S}[0,r]$ At classic level, the size of this mode space is infinite and uncountable. But quantizations will change things abruptly.

{\em The Origin of Bek-Hwk Entropy} Essentially, the entropy of all physical objects arises from the information given up by the investigator about their inner structure and motion state. For pBHs, we will show in this section that the amount of information related with their radial mass profile in the Schwarzschild time definition or the oscillatory mode in the Lema\^itre time definition has the exponentiated area law feature. So their Bek-Hwk entropy arises from the investigator's giving up of these information in believing that they have successfully implemented event horizon thus non-recognizable inner-structures. Since the classic metric families \eqref{metricOFO} and \eqref{metricICO} form a pair of equally right and complete descriptions for the microscopic state of pBHs. We choose to quantize the microscopic state of collapsing materials characterised by the $m^{\scriptscriptstyle\!S}[0,r]\otimes\dot{m}^{\scriptscriptstyle\!S}[t=0,2GM_\mathrm{tot}]$ parameters in the metric family \eqref{metricOFO} and calculate the resultant wave-functional's degeneracy. 

For this purpose, we consider the collapsar as a combination of many concentric shells $m_i$ and use $\{m_i\}$ to denote an arbitrary partition of the total mass $M_\mathrm{tot}$ \cite{dfzeng2023,dfzeng2018a,dfzeng2018b}. The effective geometry controlling each shell $m_i$'s motion can be written  as
\bea{}
&&\hspace{-5mm}ds^2=-h_idt^2+h_i^{-1}dr^2+r^2d\Omega^2
\label{effMetricmi}, h_i=1-\frac{2GM_i}{r},
\eea
where $M_i=\sum_{i'=1}^{i}m_{i'}$ is the mass of shell $i$ and its inner partners together. This shell's motion is determined by the standard geodesic equation and four velocity normalization of a representative volume element on it, which can be written into a hamiltonian constraint
\beq{}
\frac{1}{2}m_i\dot{x}^2-\frac{GM_im_i}{x}-m_i(\gamma_i^2-1)=0,
\label{constraintB}
\eeq
where $\gamma_i=h_i\dot{t}$ is an integration constant of the geodesic equation. $\gamma_i=0~\mathrm{and}~1$ correspond to cases the shell is released from $x^i_\mathrm{ini}=2GM_i$ and $x^i_\mathrm{ini}=\infty$ respectively. This equation can be quantised canonically
\beq{}
[-\frac{\hbar^2}{2m_i}\partial_x^2-\frac{GM_im_i}{x}-m_i(\gamma_i^2-1)]\psi_i(x)=0,
\label{eigenStateSchr\"odinger}
\eeq
where $\psi_i(x)$ denotes the probability amplitude the shell be measured of size $x$. All other shells can be done similarly. Directly multiplying their wave-functions together, we will get the wave functional of the whole collapsar as follows
\begin{align}
\Psi[M(r)]=\psi_0\otimes\psi_1\otimes\psi_2\cdots, \sum_im_i=M_\mathrm{tot}.
\label{directProductWaveFunction}
\end{align}
Except the normalization of $\psi_i$, equation \eqref{eigenStateSchr\"odinger} is almost the standard eigenstate Schr\"odinger equation with coulomb potentials. Its solution \cite{dfzeng2018a,dfzeng2018b,dfzeng2021,dfzeng2022} can be written down immediately
\bea{}
&&\hspace{-3mm}\psi_i=N_ie^{-x}xL_{n_i-1}^1(2x),x\equiv m_ir(1\mm\gamma_i^2)^\frac{1}{2}/\hbar
\\
&&\hspace{-3mm}n_i=\frac{GM_im_i}{\hbar(1\mm\gamma_i^2)^\frac{1}{2}}=1,2,3\cdots
\label{niRule}
\eea
where $L_{n_i-1}^1(2x)$ is the associated Lagurre polynomial and $N_i$ is the normalization of each single shell's wave-function; $n_i$ is the corresponding radial excitation level. 

The wave-functional \eqref{directProductWaveFunction} of the whole collapsar is determined by the scheme of shell $\mathbf{p}$artition $\{m_i\}$ and radial $\mathbf{e}$xcitation $\{n_i\}$. Any $\mathbf{p}\otimes\mathbf{e}$ scheme satisfying 
\bea{}
&&\hspace{-7mm}(i){\sum\nolimits_i}\!m_i{=}M_\mathrm{tot};(ii) \mathrm{the~position~}r^{|\psi^2_i|}_\mathrm{max}\mathrm{~of}|\psi_i^2|'s\mathrm{~global}
\nonumber
\\
&&\hspace{-7mm}\mathrm{maximal~value~}\mathrm{lies~outside}~r^i_h{=}2GM_i~\mathrm{for~all~}i~\mathrm{(to}
\label{pecondition}
\\
&&\hspace{-7mm}\mathrm{avoid~shells~fall~into~horizon~in~finite}~t~\mathrm{time)}
\nonumber\\
&&\hspace{-7mm}(iii)r^{|\psi^2_{i}|}_\mathrm{max}{\leqslant}r^{|\psi^2_{i+1}|}_\mathrm{max} \mathrm{for~all}~i~(\mathrm{to~avoid~repeating})
\nonumber
\eea
will lead to a possible microscopic state of the collapsar. These states can be constructed from the basic $\mathbf{p}\otimes\mathbf{e}$ scheme represented the following way,
\bea{}
&&\hspace{-7mm}M_{k~}\textcolor{cyan}{\rule[-4pt]{55mm}{4mm}\,\rule[-4pt]{5mm}{4mm}\hspace{-5mm}}m_k,n_k{=}2
\label{groundpe}
\\
&&\hspace{-7mm}M_{\hspace{-1pt}k^{\!-}}\textcolor{cyan}{\rule[-4pt]{49.5mm}{4mm}\,\rule[-4pt]{5mm}{4mm}\hspace{-5mm}}m_{\hspace{-1pt}k^{\!-}},n_{k^{\!-}}{=}2
\nonumber
\\
&&\hspace{-7mm}M_{i~}\cdots\cdots\rule{20mm}{0pt}\textcolor{cyan}{\rule[-4pt]{5mm}{4mm}\hspace{-5mm}}m_i\,,n_{i}{=}2
\nonumber
\\
&&\hspace{-7mm}M_{2~}\textcolor{cyan}{\rule[-4pt]{5mm}{4mm}\,\rule[-4pt]{5mm}{4mm}\hspace{-5mm}}m_2,n_2{=}2
\nonumber
\\
&&\hspace{-7mm}M_{1~}\textcolor{cyan}{\rule[-4pt]{5mm}{4mm}\hspace{-5mm}}m_{1},n_1{=}2
\nonumber
\eea
through the shell's recombination and the radial excitation's reassigning,
\bea{}
&&\hspace{-7mm}M_{\hspace{-1pt}q}\textcolor{cyan}{\rule[-4pt]{48mm}{4mm}\,\rule[-4pt]{9mm}{4mm}\hspace{-6mm}}m_{\hspace{-1pt}q}~,n_{q}{=}n^\mathrm{min}_{q}
\label{excitationpe}
\\
&&\hspace{-7mm}M_{i~}\cdots\cdots\rule{15mm}{0pt}\textcolor{cyan}{\rule[-4pt]{14mm}{4mm}\hspace{-8mm}}m_i~~~~,~\textcolor{red}{n_{i}{\geqslant}2}
\nonumber
\\
&&\hspace{-7mm}M_{2~}\textcolor{cyan}{\rule[-4pt]{10mm}{4mm}\,\textcolor{red}{\xrightarrow{~~}}\rule[-4pt]{9mm}{4mm}\hspace{-6mm}}m_2~~,\textcolor{red}{n_2{\geqslant}2}
\nonumber
\\
&&\hspace{-7mm}M_{1~}\textcolor{cyan}{\textcolor{red}{\xrightarrow{~~}}\rule[-4pt]{5mm}{4mm}\hspace{-5mm}}m_{1},\textcolor{red}{n_1{\geqslant}2}
\nonumber
\eea
The $\mathbf{p}\otimes\mathbf{e}$ scheme \eqref{groundpe} is called basic because in it the global maximal values of all shells' wave function happen on their minimal possible radial position allowed by the three conditions in \eqref{pecondition} so that all $\gamma_i=0$ and all $n_i{=}2$,
\beq{}
\frac{GM_im_i}{\hbar(1-0)^\frac{1}{2}}=2,\forall i{\in}\{1,2,\cdots,k\}\Rightarrow{}k{=}\frac{GM_\mathrm{tot}^2}{2\sqrt{2}\hbar}.
\eeq 
This derivation holds as long as $M_\mathrm{tot}$ is mildly larger than $M_\mathrm{pl}$\cite{dfzeng2021,dfzeng2023}. The radial excitation number $n_q^\mathrm{min}$ of the excited $\mathbf{p}\otimes\mathbf{e}$ \eqref{excitationpe} is determined by the requirement that the global maximal value happens on the minimal position $r^{|\psi_q^2|}_\mathrm{max}$ allowed by \eqref{pecondition}-(ii), i.e. $2GM_\mathrm{tot}<r^{|\psi_q^2|}_\mathrm{max}$. For each recombination scheme $\{m_i\}$, the radial excitation scheme $\{n_i\}$ is approximately unique.  So the correspondence between parameters characterizing the quantum wave-function and classic inner-structure geometry reads
\beq{}
\{m_i\}\otimes\{\prod_i^{\otimes}n_i\}\leftrightarrow m^{\scriptscriptstyle\!S}(0,r)\otimes\dot{m}(0,2GM_\mathrm{tot}).
\eeq 
 The total number of ways recombining shells in \eqref{groundpe} is exactly $2^k=\exp\{\frac{GM_\mathrm{tot}^2}{2\sqrt{2}\hbar}\log2\}$.
For certain recombined-shell partition \cite{dfzeng2023}, the radial excitation scheme $\prod_i^{\otimes}n_i$ allowed by \eqref{pecondition} have a polynomial type non-uniqueness $X(A)$. So the degeneracy of the wave-functional \eqref{directProductWaveFunction} becomes
 \beq{}
 W{=}\exp\{c(\epsilon)[A/4G\hbar{+}\ln X(A)]\},A{=}4\pi (2GM_\mathrm{tot}\!)^2\!,
 \label{areaLawFormula}
 \eeq
 where $\epsilon$ parametrises the precision of the shell partition and $c(\epsilon)$ is an $\epsilon$-dependent constant and $M_\mathrm{tot}$ is only required to be mildly larger than $G^{-\frac{1}{2}}{\equiv}M_\mathrm{pl}$. The parameter $\epsilon$ enters here because we need a standard to define to what degree two partition schemes $\{m_i\}$ and $\{m_i'\}$ satisfying conditions \eqref{pecondition} simultaneously and have equal $\{n_i\}$ are distinguishable from each other. This parameter affects the value of $k$ linearly so by setting $\epsilon=\frac{\ln2}{8\pi}$ we will get the exact Bek-Hwk entropy formula with logarithmic corrections.

Because the parameters $m^{\scriptscriptstyle\!L}[\varrho]$ and $m^{\scriptscriptstyle\!S}[0,r]$ in classic geometries are one-to-one correspondence, the above proof implies that the quantization of over-cross-oscillatory solid-balls in the Lema\^itre time definition will lead to the same results as the quantization of collapsars in the Schwarzschild time definition. So the two pictures are two equally right and complete descriptions for the microscopic state of pBHs. The Bek-Hwk entropy of BHs arises from the investigator's giving up of the information encoded by the collapsing material's radial distribution or the over-cross-oscillatory material's oscillatory mode. This is possible, because in this explanation of Bek-Hwk entropy, the carrier of the fundamental degrees of freedom is the collapsing material's collective motion mode instead of their compositing particles. According to \eqref{niRule} the typical mass of these modes is $1/GM$, so their total number is $\mathcal{O}[GM^2]$. Since this is the typical mass of Hawking particles, it is very natural to link this fundamental degrees of freedom with the information missing puzzle \cite{dfzeng2021,dfzeng2022}. In fact, this is just the road we reach the concept of CGC in this work.

{\em Potential observational signals} Our proof of the area law formula of Bek-Hwk entropy implies that, to all probes defined in the coverage of the Schwarzschild time definition, pBHs should be considered as collapsars with extended radial-mass profile and only close-to-implementing but not successfully implemented horizon. When two pBHs inspiral and merge, both are probes of their partners, so will see no horizons on each other but will exhibit mass transferring and redistributions under mutual-gravitation and radiation back-reaction. This signal should be extractable from the data of BH binary's merger processes \cite{GWTC1,GWTC2,GWTC3}. In \cite{dfzeng2025b}, we will refine an exact one body method proposed in \cite{dfzeng2023} and calculate the gravitational wave (GW) of such processes and uncover that, their late-time damping feature is possible only when the participating BHs carry extended inner structure and not successfully implemented horizon. Numeric relativity \cite{NR2004,NR2007,NR2018,NR2019} provides us right waveforms only because it replaces the mass transferring and redistribution of the participating BHs with the shape deformation of their apparent horizon.  Other signals such as echos from BHs with incomplete absorbing horizon \cite{cardoso2016, gwEchoCardoso1902,gwEchoMaggio1907,gwEchoKaloper1912,gwEchoCardoso2007,gwEchoStojkovic2011,gwEchoMaggio2202} and special features of the shadow image similar with the string theory fuzzy balls \cite{mayerson2021fuzzball,mayerson2022microReview} are also possible in our pBHs pictures. 
 
In astronomical contexts, possessing extended inner-mass profile and only close-to-implementing but not successfully implemented horizon will make pBHs Neutron Star (NS) like collapsars more than no hair objects. Differences between pBHs and NSs may only be, pBHs have continuously contracting matter core while NSs have static and balanced matter configuration; pBHs can carry magnetism even no-charges and with their magnetic dipole parallel with angular momentum while NSs can have these two quantities point to different directions. We will show in \cite{dfzeng2025c} that perturbations in the dynamically contracting plasmonic crust of this BHs caused by the inhomogeneous neutron synthesis process will oscillate as relativistic sound waves and stimulate fast radio bursts by changing the equivalent molecular current of the system. In the case of no perturbation, such BHs will not radiate as NSs because their magnetic field aligns with angular momentum. This is a new mechanism of fast radio bursts which is not allowed by other BH inner-structure pictures reviewed in the introduction part of this work. We will take it as an exclusive prediction of our inner-structure picture for BHs and discuss in depth in \cite{dfzeng2025c}.

{\em Conclusion and discussion} Our analysis here shows that GR permits two complementary description for the microscopic state of pBHs: (a) continuously contracting collapsars with close-to-implementing but never successfully implemented horizon in the Schwarzschild time definition, and (b) over-cross oscillatory solid-balls ergodically but unpredictably crossing the singularity in the Lema\^itre time definition. This complementarity emerges naturally from the general coordinate invariance of GR rather than apriori assumptions required of any unknown quantum gravitation theory. The area-law formula of Bekenstein-Hawking entropy arises from the degeneracy of the quantum wave functional of the collapsing material's configuration. This resolves the contradiction between the statistic feature of BH thermodynamics and the deterministic scenario of gravitational collapse implied by singularity theorems. It suggests that to all outside probes, pBHs should be considered compact objects inherit radial mass-profile from their progenitors rather than singular objects wrapped by successfully implemented event horizon. This is a remarkable declaration dis-verfiable through gravitational wave spectroscopy and multi-messenger astronomy.

In contrast with the thermodynamic analogue \cite{hawking1971,Bekenstein1972,Bekenstein1974} and the Euclidean path integral proof \cite{GibbonsHawking1977,BrownYork1994,entropyAshokeLogCorrec2013} based on the saddle point approximation, our proof of the area-law formula here provides an intuitive and first-principle based microscopic interpretation for the origin of Bek-Hwk entropy.  Our interpretation challenges those based on string theory and loop quantum but are fully compatible with GR, yet requires no beyond-GR mechanisms. It only adopts the inherent state-counting freedom within Einstein's theory itself. The predicted ``horizonless'' phenomenology differs crucially from the traditional BH models in three aspects: (1) the ring-down of gravitational waves following from the binary merger process contain effects from the BH inner structure's deformation; (2) electromagnetic counterparts may emit characteristic radio wave signatures due to the shell-cracking and relaxation in their early stages; (3) quantum emission processes maintain unitary evolution without information loss. The former two aspects enable direct observational tests using existing gravitational wave detectors and radio transient surveys. 

The last aspects concerns the nature of quantum gravity. By our complementarity, pBHs can be understood as super massive atoms bounded by gravitation, they will experience gravity induced spontaneous radiation just as usual atoms experience electromagnetic force induced spontaneous radiation \cite{dfzeng2021,dfzeng2022}. This radiation has exactly thermal spectrum due to the BHs' exponentially high degree of microscopic state degeneracy. It allows for explicitly hermitian hamiltonian description thus no information missing in any sense, see \cite{dfzeng2023} for analysis related with the fire wall paradoxes. The analysis there implies that, considering the complementarity of gravitational collapse, the contradiction between GR and quantum mechanics will get remarkable reduction. This is obviously a positive news for their unification \cite{quanGravity}.

{\bfseries Acknowledgements} We thank very much to Asta Heinesen, Rico Lo, Rodrigo Panosso, Takuya Katagiri, Hai Lin and Zhengwen Liu for their listening and advices. Part of this work is done at Niels Bohr Institute and gets sponsors from Chinese Scholarship Council for oversea visiting research under grant no. CSC202006545026, Vitor Cardoso and Julie de Molade provide warm hosts at NBI. This work is supported by the NSFC grant no. {11875082}, 
the Villum Investigator program of VILLUM Foundation (grant no. VIL37766) and the DNRF Chair program (grant no. DNRF162) by the Danish National Research Foundation and the European Union's Horizon 2020 research and innovation programme under the Marie Sklodowska-Currie grant No. 101131233.


\begin{thebibliography}{00}

\bibitem{Penrose1965}
R. Penrose, 
``Gravitational Collapse and Space-Time Singularities'',
{\it Phys. Rev. Lett.} {\bf14 (3)}, 57(1965)

\bibitem{Hawking1976}
S. Hawking,
``Breakdown of Predictability in Gravitational Collapse'',
{\em Phys. Rev. } {\bf D14, } 2460 (1976)

\bibitem{geroch1979}
R. Geroch and G. Horowitz, 
``Global structure of spacetimes, in General Relativity: An Einstein Centenary Survey'', 
{\bf pp. 212–293}, 1979.

\bibitem{Kerr2023}
R. P. Kerr,
``Do Black Holes have Singularities?''
\href{https://doi.org/10.48550/arXiv.2312.00841}{doi:10.48550/arXiv.2312.00841}

\bibitem{CCH1969}
R. Penrose,
``Gravitational collapse: The role of general relativity'',
{\em Riv. Nuovo Cim.} {\bf 1} (1969) 252, {\em Gen. Rel. Grav.} {\bf 34} (2002) 1141,
\href{https://link.springer.com/article/10.1023/A:1016578408204}{DOI: 10.1023/A:1016578408204}.

\bibitem{CCH1979}
R. Penrose,
``Singularities and time-asymmetry'',
Penrose, R. Israel, W. (Ed.). (1979). Singularities and time-asymmetry. United Kingdom: University Press.

\bibitem{strominger1996}
A. Strominger, C. Vafa,
``Microscopic Origin of the Bekenstein-Hawking Entropy'',
{\it Phys. Lett.} {\bf B379} (1996)99,
\href{http://arxiv.org/abs/hep-th/9601029}{arXiv:hep-th/9601029}.

\bibitem{HorowitzStrominger9602}
G. Horowitz, A. Strominger,
``Counting States of Near-Extremal Black Holes'',
{\em Phys. Rev. Lett.} {\bf 77} (1996) 2368-2371,
\href{https://doi.org/10.1103/PhysRevLett.77.2368}{doi.org/10.1103/PhysRevLett.77.2368}

\bibitem{fuzzballsMathur2002}
O. Lunin and S. D. Mathur,
``Statistical Interpretation of the Bekenstein Entropy for Systems with a Stretched Horizon'',
{\it Phys. Rev. Lett.} {\bf 88}(2002) 211303,
\href{https://doi.org/10.1103/PhysRevLett.88.211303}{DOI:10.1103/PhysRevLett.88.211303}

\bibitem{fuzzballSkenderis2007}
K. Skenderis and M. Taylor,
``Fuzzball Solutions for BHs and D1-Brane–D5-Brane Microstates'',
{\it Phys. Rev. Lett.} {\bf 98} (2007) 071601,
\href{https://doi.org/10.1103/PhysRevLett.98.071601}{DOI:10.1103/PhysRevLett.98.071601}.

\bibitem{lqgEntropy1996}
C. Rovelli,
``Black hole entropy from loop quantum gravity'',
{\it Phys. Rev. Lett.} {\bf 77} (1996) 3288,
\href{http://arxiv.org/abs/gr-qc/9603063}{arXiv:gr-qc/9603063}.

\bibitem{lqgEntropy1997}
A. Ashtekar, J. Baez, A. Corichi, K. Krasnov,
``Quantum Geometry and Black Hole Entropy'',
{\it Phys. Rev. Lett.} {\bf80} (1998) 904,
\href{http://arxiv.org/abs/gr-qc/9710007}{arXiv:gr-qc/9710007}.

\bibitem{AshtekarBojowald0509}
A. Ashtekar and M. Bojowald, 
``{Quantum geometry and the Schwarzschildsingularity}'',
 {\em Classical and Quantum Gravity} {\bf 23} (2006) 391--411,
\href{https://arxiv.org/abs/gr-qc/0509075}{arXiv:gr-qc/0509075}.

\bibitem{Modesto0509}
L. Modesto, ``{Loop quantum black hole}'', 
{\em Classical and Quantum Gravity}
{\bf 23} (2006) 5587--5601,
\href{https://arxiv.org/abs/gr-qc/0509078}{arXiv:gr-qc/0509078}.

\bibitem{softhairHPS2016}
S. Hawking, M. Perry, A. Strominger,
``Soft Hair on Black Holes'',
{\em Phys.Rev.Lett.} {\bf 116} (2016) 231301,
\href{https://doi.org/10.1103/PhysRevLett.116.231301}{DOI: 10.1103/PhysRevLett.116.231301},
\href{https://arxiv.org/abs/1601.00921}{arXiv:hep-th/1601.00921}

\bibitem{vijay2024prl}
V. Balasubramanian, A. Lawrence, J. Magan, M. Sasieta,
``Microscopic origin of the entropy of astrophysical black holes'',
{\em Phys.Rev.Lett} {\bf 132} (2024) 141501,
\href{https://doi.org/10.1103/PhysRevLett.132.141501}{DOI: 10.1103/PhysRevLett.132.141501},
\href{https://arxiv.org/abs/2212.08623}{arXiv:hep-th/2212.08623}.

\bibitem{dfzeng2023}
Ding-fang Zeng, 
``Microscopic State of BHs and an Exact One Body Method for Binary Dynamics in General Relativity'',
{\em EPJC} {\bf 84} (2024) 370,
\href{https://doi.org/10.48550/arXiv.2311.11764}{arXiv: 2311.11764}

\bibitem{complementarity1990thooft}
G. 't Hooft,
``The BH Interpretation of String Theory'',
{\em Nucl. Phys.} {\bf B335} (1990) 138,
\href{https://doi.org/10.1016/0550-3213(90)90174-C}{doi.org/10.1016/0550-3213(90)90174-C}

\bibitem{complementarity1993schoutens}
K. Schoutens, E. Verlinde, H. Verlinde
``Quantum BH Evaporation'',
{\em Phys.Rev.} {\bf D48} (1993) 2670,
\href{https://doi.org/10.1103/PhysRevD.48.2670}{doi.org/10.1103/PhysRevD.48.2670}

\bibitem{complementarity1993}
L. Susskind, L. Thorlacius, and J. Uglum,
``The stretched horizon and BH complementarity''
{\em Phys. Rev.} {\bf D48} (1993) 3743,
\href{https://doi.org/10.1103/PhysRevD.48.3743}{doi.org/10.1103/PhysRevD.48.3743}

\bibitem{dfzeng2017}
Ding-fang Zeng, 
``Resolving the Schwarzschild singularity in both classic and quantum gravities'',
{\em Nucl. Phys.} {\bf B917} (2017) 178, arXiv: 1606.06178,
\href{https://doi.org/10.1016/j.nuclphysb.2017.02.005}{doi.org/10.1016/j.nuclphysb.2017.02.005}.

\bibitem{dfzeng2020} 
Ding-fang Zeng, 
``Exact inner metric and microscopic state of AdS$_3$-Schwarzschild BHs'',
{\em Nucl. Phys.} {\bf B954} (2020) 115001, arXiv: 1812.06777,
\href{https://doi.org/10.1016/j.nuclphysb.2020.115001}{doi.org/10.1016/j.nuclphysb.2020.115001}.

\bibitem{dfzeng2021}
Ding-fang Zeng, 
``Spontaneous Radiation of BHs'',
{\em Nucl. Phys.} {\bf B977} (2022) 115722, arXiv: 2112.12531,
\href{https://doi.org/10.1016/j.nuclphysb.2022.115722}{doi.org/10.1016/j.nuclphysb.2022.115722}

\bibitem{dfzeng2022}
Ding-fang Zeng, 
``Gravity Induced Spontaneous Radiation'',
{\em Nucl. Phys.} {\bf B990} (2023) 116171,arXiv: 2207.05158,
\href{https://doi.org/10.1016/j.nuclphysb.2023.116171}{doi.org/10.1016/j.nuclphysb.2023.116171}

\bibitem{dfzeng2018a}
Ding-fang Zeng, 
``Schwarzschild Fuzzball and Explicitly Unitary Hawking Radiations'',
{\em Nucl. Phys.} {\bf B930} (2018) 533-544, arXiv: 1802.00675,
\href{https://doi.org/10.1016/j.nuclphysb.2018.03.012}{doi.org/10.1016/j.nuclphysb.2018.03.012}.

\bibitem{dfzeng2018b}
Ding-fang Zeng, 
``Information missing puzzle, where is hawking's error?'',
{\em Nucl. Phys.} {\bf B941} (2018) 665, arXiv: 1804.06726,
\href{https://doi.org/10.1016/j.nuclphysb.2019.02.023}{doi.org/10.1016/j.nuclphysb.2019.02.023}.

\bibitem{Stojkovic2008b}
E. Greenwood, D. Stojkovic,
``Quantum gravitational collapse: non-singularity and non-locality''
{\em JHEP} {\bf0806} (2008) 042,
\href{https://arxiv.org/abs/0802.4087}{0802.4087}

\bibitem{Stojkovic2014}
A. Saini, D. Stojkovic,
``Non-local (but also non-singular) physics at the last stages of gravitational collapse'',
{\em Phys. Rev.} {\bf D89} (2014) 044003
\href{https://arxiv.org/abs/1401.6182}{1401.6182}

\bibitem{dfzeng2025b}
Ding-fang Zeng, 
``Inspiral to Merger, a Full Process Applicable One Body Method to the Two Body problem in General Relativity'',
to appear as the second cousin of this manuscript. 

\bibitem{dfzeng2025c}
Ding-fang Zeng, 
``Complementarity inside BHs, Fast Radial Burst as a Prediction'',
to appear as the third cousin of this manuscript. 

\bibitem{CPTsymmUniverse2018}
L. Boyle, K. Finn, and N. Turok,
``CPT-Symmetric Universe'',
{\it Phys. Rev. Lett.} {\bf 121} (2018) 251301,
\href{https://doi.org/10.1103/PhysRevLett.121.251301}{DOI:10.1103/PhysRevLett.121.251301}

\bibitem{CPTsymmUniverse2022}
L. Boyle, K. Finn, N. Turok,
``The Big Bang, CPT, and neutrino dark matter'',
{\it Annals of Physics} {\bf 438} (2022) 168767,
\href{https://doi.org/10.1016/j.aop.2022.168767}{DOI:10.1016/j.aop.2022.168767}

\bibitem{GWTC1}
``GWTC-1: A Gravitational-Wave Transient Catalog of Compact Binary Mergers Observed by LIGO and Virgo during the First and Second Observing Runs'', 
GWTC-1: {\em Phys. Rev.} {\bf X9} (2019) 031040, arXiv: 1811.12907,
\href{https://doi.org/10.1103/PhysRevX.9.031040}{doi.org/10.1103/PhysRevX.9.031040}.

\bibitem{GWTC2}
``GWTC-2: Compact Binary Coalescences Observed by LIGO and Virgo During the First Half of the Third Observing Run'',
GWTC-2: {\em Phys. Rev.} {\bf X11} (2021) 021053,
\href{https://doi.org/10.1103/PhysRevX.11.021053}{doi.org/10.1103/PhysRevX.11.021053}.

\bibitem{GWTC3}
``GWTC-3: Compact Binary Coalescences Observed by LIGO and Virgo During the Second Part of the Third Observing Run'',
GWTC-3: \href{https://doi.org/10.48550/arXiv.2111.03606}{doi.org/10.48550/arXiv.2111.03606}.

\bibitem{NR2004}
M. Ansorg, B. Brugmann, W. Tichy, 
``Single-domain spectral method for black hole puncture data'', 
{\em Phys. Rev.} {\bf D70} (2004) 064011, arXiv: gr-qc/0404056
\href{https://doi.org/10.1103/PhysRevD.70.064011}{doi.org/10.1103/PhysRevD.70.064011}

\bibitem{NR2007}
Z. B. Etienne, J. A. Faber, Y. T. Liu et al, 
``Filling the holes: Evolving excised binary black hole initial data with puncture techniques'', 
{\em Phys. Rev.} {\bf D76} (2007) 101503,
\href{https://doi.org/10.1103/PhysRevD.76.101503}{doi.org/10.1103/PhysRevD.76.101503}.

\bibitem{NR2018}
V. Varma, M. A. Scheel,H. P. Pfeiffer, 
``Comparison of binary black hole initial data sets'', 
{\em Phys. Rev.} {\bf D98} (2018) 104011, arXiv: 1808.08228,
\href{https://doi.org/10.1103/PhysRevD.98.104011}{doi.org/10.1103/PhysRevD.98.104011}.

\bibitem{NR2019}
D. Pook-Kolb, O. Birnholtz, B. Krishnan, and E. Schnetter,
``Interior of a Binary Black Hole Merger'',
{\em Phys. Rev. Lett.} {\bfseries 123} (2019) 171102, arXiv:1903.05626.
\href{https://doi.org/10.1103/PhysRevLett.123.171102}{doi.org/10.1103/PhysRevLett.123.171102}

\bibitem{cardoso2016}
V. Cardoso, S. Hopper, C F. B. Macedo and et al,
``Gravitational-wave signatures of exotic compact objects and of quantum corrections at the horizon scale'',
{\it Phys. Rev.} {\bf D94} (2016) no.8, 084031,
\href{https://arxiv.org/abs/1608.08637}{arXiv: 1608.08637}.

\bibitem{gwEchoCardoso1902} 
V. Cardoso, V. F. Foit, M. Kleban,
``Gravitational wave echoes from black hole area quantization'',
{\em JCAP} {\bf 08} (2019) 006, arXiv: 1902.10164,
\href{https://doi.org/10.1088/1475-7516/2019/08/006}{doi.org/10.1088/1475-7516/2019/08/006}.

\bibitem{gwEchoMaggio1907}
E. Maggio, A. Testa, S. Bhagwat, P. Pani,
``Analytical model for gravitational-wave echoes from spinning remnants'',
{\em 	Phys. Rev.} {\bf D100} (2019) 064056 (2019), arXiv: 1907.03091,
\href{https://doi.org/10.1103/PhysRevD.100.064056}{doi.org/10.1103/PhysRevD.100.064056}

\bibitem{gwEchoKaloper1912} 
G. D’Amico, N. Kaloper, 
 ``On Black Hole Echoes'', 
 {\em Phys. Rev.} {\bf D102} (2020) 044001, arXiv: 1912.05584,
 \href{https://doi.org/10.1103/PhysRevD.102.044001}{doi.org/10.1103/PhysRevD.102.044001}.
 
\bibitem{gwEchoCardoso2007} 
I. Agullo, V. Cardoso, A. del Rio etal, 
``Potential gravitational-wave signatures of quantum gravity'', 
{\em Phys. Rev. Lett.} {\bf 126} (2021) 041302, arXiv: 2007.13761,
\href{https://doi.org/10.1103/PhysRevLett.126.041302}{doi.org/10.1103/PhysRevLett.126.041302}.

\bibitem{gwEchoStojkovic2011}
R.f. Dong, D. Stojkovic, 
``Gravitational wave echoes from black holes in massive gravity'',
{\em Phys. Rev.} {\bf D103} (2021) 026058, arXiv: 2011.04032,
\href{https://doi.org/10.1103/PhysRevD.103.024058}{doi.org/10.1103/PhysRevD.103.024058}.

\bibitem{gwEchoMaggio2202}
S. Chakraborty, E. Maggio, A. Mazumdar, P. Pani,
``Implications of the quantum nature of the black hole horizon on the gravitational-wave ring-down'',
{\em Phys. Rev.} {\bf D106} (2022) 024041, arXiv: 2202.09111.
\href{https://doi.org/10.1103/PhysRevD.106.024041}{doi.org/10.1103/PhysRevD.106.024041}.

\bibitem{mayerson2021fuzzball}
F. Bacchini, D. R. Mayerson, B. Ripperda et al,
``Fuzzball Shadows: Emergent Horizons from Microstructure'',
{\it Phys. Rev. Lett.} {\bf127} (2021) 171601 

\bibitem{mayerson2022microReview}
D. R. Mayerson,
``Modave Lectures on Horizon-Size Microstructure, Fuzzballs and Observations'',
\href{https://arxiv.org/abs/2202.11394}{arxiv:2202.11394}

\bibitem{hawking1971}
S. W. Hawking,
``Gravitational Radiation from Colliding Black Holes'',
\href{https://doi.org/10.1103/PhysRevLett.26.1344}{Phys. Rev. Lett. 26, 1344}

\bibitem{Bekenstein1972}
J. D. Bekenstein,
``Black Holes and Entropy'',
\href{https://doi.org/10.1103/PhysRevD.7.2333}{Phys. Rev. D 7, 2333}

\bibitem{Bekenstein1974}
J. D. Bekenstein
``Generalized second law of thermodynamics in black-hole physics'',
\href{https://doi.org/10.1103/PhysRevD.9.3292}{Phys. Rev. D 9, 3292}

\bibitem{entropyAshokeLogCorrec2013}
A. Sen,
``Logarithmic corrections to Schwarzschild and other non-extremal BH entropy in different dimensions'',
\href{https://doi.org/10.1007/JHEP04(2013)156}{DOI:10.1007/JHEP04(2013)156}

\bibitem{GibbonsHawking1977}
G. W. Gibbons and S. W. Hawking,
``Action integrals and partition functions in quantum gravity'',
{\em Phys. Rev.} {\bf D15} (1977) 2752,
\href{https://doi.org/10.1103/PhysRevD.15.2752}{doi.org/10.1103/PhysRevD.15.2752}.

\bibitem{BrownYork1994}
J.D. Brown, J.W. York,
``The Path Integral Formulation of Gravitational Thermodynamics'', reports on the conference "The Black Hole 25 Years After", Santiago, Chile, January 1994
\href{https://doi.org/10.48550/arXiv.gr-qc/9405024}{doi.org/10.48550/arXiv.gr-qc/9405024}

\bibitem{quanGravity}
G. 't Hooft,
``Conversations on Quantum Gravity'',
J. Armas (Ed.) (2021), pp275.  Cambridge Press.

\end{thebibliography}
\end{document}